# Smectic ordering in athermal systems of rod-like triblock copolymers

Szabolcs Varga[a)] and Seth Fraden[b)]

[a)] *Department of Physics, University of Veszprém, H-8200 Veszprém, PO Box 158, Hungary*

[b)] *Complex Fluids Group, Martin Fisher School of Physics, Brandeis University, Massachusetts 02454 USA*

Number of pages: 22 (including figure captions and figures)

[a)] e-mail: vargasz@almos.vein.hu

[b)] e-mail: fraden@brandeis.edu

**Abstract**

The phase behavior of the system of parallel rigid triblock copolymers is examined using the second-virial density functional theory. The triblock particle consists of two identical infinitely thin hard rods of finite lengths on the opposite ends of one central hard cylinder with nonzero length and diameter. Stability analyses and free energy calculations show that the system of parallel particles can form not only uniform nematic and smectic A phases, but a smectic-C phase too. The stability and structure of the tilted structure is controlled by only the diameter and the length of the central cylinder segment. Interestingly, the diameter effects only the layer tilting and the periodicity, but not the packing fraction of the nematic to smectic-C transition. For all values of cylinder length the usual smectic-A and smectic-C transitions compete with each other and no nematic-columnar transition is observed. At low and high cylinder length the smectic-A phase is stabilized first, while the smectic C is the most stable for intermediate length values.

## 1. Introduction

The theory of liquid crystalline order based arising from hard core repulsion of anisotropic shaped molecules begins with Onsager[1] who explained the isotropic – nematic phase transition for colloidal suspensions of rods and plates. Because the interparticle interaction is purely steric, only entropy enters the free energy of a fluid of hard particles. Interparticle potentials are composed of repulsive and attractive components and in the molecular theory of liquids the free energy can be treated as primarily arising from the repulsive component with attractions added as a perturbation[2]. For this reason a line of inquiry that has occupied many after Onsager has been to explore what other liquid crystalline phase transitions could be understood as arising for purely entropic reasons and what phase transitions required attractions[3].

Entropic theories of the nematic to smectic – A of hard core rodlike particles were first treated by Kimura and co-workers[4] and verified in simulations by Stroobants *et al* [5]. This latter work stimulated a large number of papers on the theory, simulation and experiment of hard core liquid crystals[6].

In this paper we are concerned with three phase transitions; the nematic to smectic – A, nematic to smectic – C, and smectic - A to smectic – C transition in hard core systems. The latter is a phase transition between a layered phase (Smectic – A) in which the long axes of the anisotropic molecules are parallel to the layer normal and a layered phase where the axes of the particles are tilted with respect to the layer (Smectic – C) as illustrated in Figure 1. This transition was first theoretically studied by Wulf[7] who considered the tilting to arise from the zig-zag shape of the mesogens, a viewpoint that was recently investigated using simulations[8]. Obliquely shaped molecules have also been treated [4,9]. Another class of molecules exhibiting the entropically driven smectic – A to smectic – C transition are rod-coil molecules, which consist of at least two blocks; one of a polymeric and one of a rigid nature [10,11]. A related molecular class consists of molecules composed of a rod grafted to a sphere, or other wider particle[12,13]. In both classes, the rodlike part drives the isotropic to nematic transition while the polymeric (spherical) part drives the smectic ordering. If the diameter of the spherical part is somewhat larger than the rod diameter then a smectic-A to smectic–C transition ensues. This class of liquid crystalline molecules is similar to block co-polymers in that the layered smectic phases can be viewed as a microphase separation arising from the frustrated macrophase separation that would occur, but is prevented from doing so by the bond between the two blocks.

Our model consisting of a cylindrical core with two thin rods extruding from the centers of the end faces (Figure 1) is a hybrid between the hard core uniform cylinders that form smectic - A and not smectic - C phases, and the rod-coil molecules, which lack inversion symmetry and exhibit bilayer phases such as the Smectic – $A_d$ and – $C_d$.[11,12] Because of the higher symmetry of the triblock cylinders our system allows study of the Smectic – A to Smectic – C transition without intervening phases.

## 2.   Theory

According to the second virial theory the free energy of an inhomogeneous system can be written as a sum of ideal and excess contributions $\beta F = \beta F_{id} + \beta F_{ex}$ [14], where

$$\beta F_{id} = \int d\underline{r} \rho(\underline{r})[\ln \rho(\underline{r}) - 1], \qquad (1)$$

$$\beta F_{ex} = -\frac{1}{2} \int d\underline{r}_1 \rho(\underline{r}_1) \int d\underline{r}_2 \rho(\underline{r}_2) f_M(\underline{r}_{12}). \qquad (2)$$

In these equations $\beta = \dfrac{1}{k_B T}$ ($k_B$ being the Boltzmann constant and $T$ the temperature), $\rho(\underline{r})$ is the local number density and $f_M(\underline{r}_{12})$ (Mayer function) is directly connected to the pair potential ($u$) through the relation of $f_M(\underline{r}_{12}) = \exp(-\beta u(\underline{r}_{12})) - 1$. Note that Eqs. (1-2) are only valid for systems without orientational freedom. In our case the pair potential is hard, so the Mayer function takes the following simple form

$$f_M(\underline{r}) = \begin{cases} -1 & \text{if } r \leq \sigma \\ 0 & \text{otherwise} \end{cases}, \qquad (3)$$

where $\sigma$ is the distance of closest approach between two particles. Since the particles are parallel and rod-like, it is reasonable to consider the stability of uniform phase (nematic) with respect to smectic-type density modulations. To do this we assume that a density modulation with a period of $g$ takes place along the $z$ axis and that the local density is uniform in a plane perpendicular to the direction of density modulation, i.e. $\rho(\underline{r}) = \rho(z) = \rho(z+g)$. Inserting the periodic local density into the free energy expression we obtain

$$\beta F / V = \frac{1}{g}\int_0^g dz \rho(z)[\ln \rho(z) - 1] + \frac{1}{2g}\int_0^g dz_1 \rho(z_1) \int_{\substack{\text{overlap}\\ \text{range of } z_{12}}} dz_2 \rho(z_2) A_{exc}(z_{12}), \qquad (4)$$

where $V$ is the volume of the system and $A_{exc}$ excluded area comes from the integration of the Mayer function in the plane normal to the density modulation. To determine the equilibrium profile of the local density we first factorize it $\rho(z) = \rho f(z)$ and then use the well-known fourier expansion method as follows

$$f(z) = 1 + \sum_{i=1}^{n} S_i \cos(ikz) \qquad (5)$$

where $\rho$ is the number density, $f(z)$ is the positional distribution function, $k$ is the wave number defined by $k = \frac{2\pi}{g}$ and $S_1 \ldots S_n$ are the expansion coefficients. After substitution of the local density (Eq. 5) into the free energy density (Eq. (4)) and using $\varphi = kz$ as a reduced variable, the ideal free energy term reduces to

$$\beta F_{id} / V = \rho \ln \rho - \rho + \rho \sigma[f], \qquad (6)$$

where

$$\sigma[f] = \frac{1}{2\pi}\int_0^{2\pi} d\varphi f(\varphi) \ln f(\varphi). \qquad (7)$$

Note that the functional $\sigma$ depends only on the expansion coefficients, but not the wave number. $\sigma$ is zero in the uniform phase, while it is positive and increases as the density profiles become more peaked in the smectic phase. This means that Eq. (7) favors the uniform density arrangement of the nematic. The excess free energy contribution to Eq. (4) becomes

$$\beta F_{ex}/V = \frac{1}{2}\rho^2 V_{exc} + \frac{\rho^2}{4}\sum_{i=1}^{n} S_i^2 \int_{\substack{overlap \\ range\ of\ z_{12}}} dz_{12}\cos(ikz_{12})A_{exc}(z_{12}), \tag{8}$$

where $V_{exc}$ is the excluded volume of a triblock particle given by $V_{exc} = \frac{D^2\pi}{4}L\left(1+7\frac{\Delta}{L}\right)$. The first term in Eq. (8) is the excess free energy of homogeneous phase, while the second part is responsible for smectic-like ordering via the minimization of the excluded volume. The total free energy now becomes

$$\beta F/V = \rho\ln\rho - \rho + \rho\sigma[f] + \frac{1}{2}\rho^2 V_{exc} + \frac{\rho^2}{4}\sum_{i=1}^{n} S_i^2 \int_{\substack{overlap \\ range\ of\ z_{12}}} dz_{12}\cos(ikz_{12})A_{exc}(z_{12}) \tag{9}$$

The excluded area of a triblock particle with total length of $L$, diameter $D$ and cylinder length $\Delta$ can be determined simply in a coordinate system, $x'y'z'$, where the $z'$ axis is parallel to the direction of the long symmetry axis of the triblock particle. It can be shown that

$$A_{exc}(z'_{12}) = \begin{cases} D^2\pi/4, & -\frac{L+\Delta}{2} < z'_{12} < -\Delta \\ D^2\pi, & -\Delta < z'_{12} < \Delta \\ D^2\pi/4, & \Delta < z'_{12} < \frac{L+\Delta}{2} \end{cases}. \tag{10}$$

The $x'y'z'$ coordinate system is applicable only if the density modulation takes place along $z'$ axis. However, the density modulation may favor a different direction to minimize most efficiently the hard body excluded region (or maximize the free volume available for the particles). In this case we encounter a tilted smectic (smectic-C) structure, where it is customary to introduce the tilt angle, $\alpha$, which is the angle between the direction of density modulation and the particle's long axis. To maintain the original free energy formalism the excluded area must be recalculated in a rotated frame, denoted the $xyz$ coordinate system, where the angle between $z$ and $z'$ axis is $\alpha$. The connection between the two coordinate systems can be expressed through the rotation transformations of $x'=x$, $y'=\cos\alpha\, y - \sin\alpha\, z$ and $z'=\sin\alpha\, y + \cos\alpha\, z$. Straightforward, but tedious calculation gives the excluded area in the rotated frame, which now depends on the tilt angle, too. In our calculations we need the Fourier components of the excluded area (see Eq. 9). The zeroth order term of the excluded area is always $V_{exc}$ irrespective of the value of the tilt angle, while the higher order terms are given by

$$\int_{\substack{overlap \\ range\,of\,z}} dz \cos(ikz) A_{exc}(z,\alpha) = \pi D^2 \left\{ \frac{J_1\left(i\frac{x_1}{2}\right)}{i\frac{x_1}{2}} \frac{\sin\left(i\frac{L+\Delta}{2}x_2\right) - \sin(i\Delta x_2)}{ix_2} + 4 \frac{J_1(ix_1)}{ix_1} \frac{\sin(i\Delta x_2)}{ix_2} \right\}, \quad i \geq 1 \tag{11}$$

where $J_1(x)$ is the first order Bessel function of the first kind, while $x_1 = k\sin(\alpha)D$ and $x_2 = k\cos(\alpha)$ are new variables. In this way the free energy of the system can be written as

$$\beta F/V = \rho \ln \rho - \rho + \rho \sigma[f] + \frac{1}{2}\rho^2 V_{exc}$$
$$+ \frac{\pi D^2 \rho^2}{4} \sum_{i=1}^{n} S_i^2 \left\{ \frac{J_1\left(i\frac{x_1}{2}\right)}{i\frac{x_1}{2}} \frac{\sin\left(i\frac{L+\Delta}{2}x_2\right) - \sin(i\Delta x_2)}{ix_2} + 4 \frac{J_1(ix_1)}{ix_1} \frac{\sin(i\Delta x_2)}{ix_2} \right\}. \tag{12}$$

In principle, the equilibrium structure and the free energy of an tilted smectic phase can be obtained by means of minimization of the above equation with respect to the wave number ($k$), the tilt angle ($\alpha$) and the fourier coefficients ($S_1 \ldots S_n$), i.e. $\frac{\partial \beta F/V}{\partial k} = 0$, $\frac{\partial \beta F/V}{\partial \alpha} = 0$ and $\frac{\partial \beta F/V}{\partial S_i} = 0$ ($i=1\ldots n$). However, the wave number and the tilt angle are coupled in our case and it is convenient to minimize with respect to $x_1$ and $x_2$ instead of $k$ and $\alpha$. To gain insight into the phase behavior of triblock particles we consider the effect of a very weak density perturbation on the free energy of the form $f(z) = 1 + S_i \cos(ikz)$ with the $S_i$ close to zero. In this case the free energy can be expanded as a function of $S_i$ and can be written as $F|_{Sm} = F|_N + aS_i^2$ in second order. The density at which the free energy of the perturbed ($F|_{Sm}$) and the homogeneous $F|_N$ phases are the same is called the bifurcation density and it is given by $a = 0$. As a result the corresponding bifurcation equation is

$$1 + \rho_{bif} \pi D^2 \left\{ \frac{J_1\left(i\frac{x_1}{2}\right)}{i\frac{x_1}{2}} \frac{\sin\left(i\frac{L+\Delta}{2}x_2\right) - \sin(i\Delta x_2)}{ix_2} + 4\frac{J_1(ix_1)}{ix_1}\frac{\sin(i\Delta x_2)}{ix_2} \right\} = 0. \qquad (13)$$

The values $x_j$ ($j$=1,2) at the bifurcation point can be obtained from $\frac{\partial a}{\partial x_j} = 0$ which is required to fulfill the minimization condition on the free energy. It turns out that the lowest density solution which corresponds to the bifurcation density is obtained with the lowest order density modulation $f(z) = 1 + S_1 \cos(kz)$, i.e. $i$=1 in all studied cases.

Before presenting the results we render the quantities dimensionless by assigning the length of the triblock particle as the unit to measure all distances. In this way our dimensionless parameters are defined and denoted with asterisks as follows: $\Delta^* = \Delta/L$, $D^* = D/L$, $k^* = kL$, $g^* = g/L$, $x_1^* = x_1$ and $x_2^* = k^* \cos(\alpha)$. We cannot use the packing fraction to make the density dimensionless because the system of infinitely thin triblock particles undergoes a nematic to smectic-A phase transition at zero packing fraction due to zero particle volume[12]. While the volume of an infinitely thin particle is zero, the excluded volume is not and therefore we use an Onsager-type dimensionless density $c = B_2^{ref} \rho$, where the $B_2^{ref}$ is the second virial coefficient of the system of infinitely thin triblock particles given by $B_2^{ref} = \frac{\pi L D^2}{8}$. We note that the relation between the packing fraction and the dimensionless density is very simple, $\eta = 2\Delta^* c$. Now we can define our dimensionless free energy

$$f^* = c\ln c - c + c\sigma[f]$$
$$+ c^2 \left( 1 + 7\Delta^* + \sum_{i=1}^{n} S_i^2 \left\{ \frac{J_1\left(i\frac{x_1^*}{2}\right)}{i\frac{x_1^*}{2}} \frac{\sin\left(i\frac{1+\Delta^*}{2}x_2^*\right) - \sin(i\Delta^* x_2^*)}{ix_2^*} + 4\frac{J_1(ix_1^*)}{ix_1^*}\frac{\sin(i\Delta^* x_2^*)}{ix_2^*} \right\} \right) \qquad (14)$$

where $f^* = \frac{\beta F B_2^{ref}}{V} + c \ln B_2^{ref}$. Note that the term $c \ln B_2^{ref}$ in the free energy does not effect the phase behavior of the system. It is an interesting feature of Eq. (14) that the $D$ dependence of the free energy is completely embedded into the reduced density and $x_1$ parameters. This means that the free energy density depends only on $\Delta^*$ at given reduced density, because the other parameters such as the fourier coefficients $S_i$ and $x_i^*$ are the results of free energy minimization. The equilibrium values of $x_1^*$ and $x_2^*$ are used to express the tilt angle ($\alpha$) and the wave number as

$$\alpha = \arctan\left(\frac{x_1^*}{x_2^* D^*}\right) \text{ and } k^* = \sqrt{\frac{x_1^{*2}}{D^{*2}} + x_2^{*2}}. \tag{15}$$

## 3. Results and discussions

First we present the results of the bifurcation analysis for the system of triblock particles with zero cylinder length ($\Delta^* = 0$) in the reduced density-tilt angle planes. This representation is advantageous because it shows the bifurcation densities of three different types of phase transitions. The first limit of $\alpha = 0°$ corresponds to the nematic-smectic A transition, the second limit of $\alpha = 90°$ is for the nematic-columnar transition, while for intermediate values of $\alpha$ a nematic-smectic C transition can exist. Figure 2 demonstrates that the bifurcation density takes its lowest value at zero tilt angle for all values of the diameter of the centre unit. This means that the system of parallel triblock particles undergoes a phase transition from nematic (uniform) phase to a smectic A phase upon compression and no nematic-smectic C or nematic-columnar transitions take place in the zero cylinder length limit. The nematic-smectic A transition is found to be second order at $c_{bif} \approx 2.3$ and $k_{bif}^* \approx 8.99$ in the same way as in the system of hard cylinders[14]. The effect of increasing length of the central unit on the bifurcation density is demonstrated in Fig. 3. The formation of the smectic C is characterized by having the minimum bifurcation density occur at a nonzero tilt angle, which first occurs at approximately $\Delta^* = 0.007$. Moreover, the tilt angle of the bifurcation point increases continuously with the cylinder length, $\Delta^*$. In addition the nematic-smectic C transition becomes more stable as the difference between the bifurcation densities of smectic C and smectic A formation increases. The effect of decreasing diameter is highlighted in Figure 4 for cylinders of length $\Delta^* = 0.02$. In this case the bifurcation density of nematic-columnar ordering is lower than that of the nematic-smectic A, but both of them are unstable as the nematic-smectic C transition has the lowest bifurcation density of $c_{bif} \approx 3.11$ irrespective of the cylinder's diameter. As a result there is a second order phase transition between the nematic and smectic C phases with the transition density being identical to the bifurcation density. The reason why the nematic-smectic C transition takes place at $c_{bif} \approx 3.11$ for all values of $D^*$ is that the dimensionless free energy and bifurcation equations (Eq. 13-14) are only indirectly related to $D^*$ through Eq. 15 as the values of $x_1^*$ and $x_2^*$ are always the same at the bifurcation point. Hence $D^*$ can only effect the tilt angle and the wave number of smectic C phase. In accordance with Eq. 15 it can be seen in Fig. 4 that the tilt angle of the smectic C phase at the bifurcation point increases with decreasing cylinder diameter ($D^*$). The results of the bifurcation analysis are summarized in Figure 5, where the reduced bifurcation densities of nematic-smectic A, nematic-smectic C and nematic-columnar transitions and the tilt angle of the possible nematic-smectic C transition are shown as a function of the cylinder's length only as the cylinder's diameter does not effect the reduced bifurcation density. Five different regimes can be identified in $c_{bif} - \Delta^*$ plane. In the first regime of $0 \leq \Delta^* \leq 0.007$ the system undergoes a nematic-smectic A transition due to the low excluded volume of the central cylinder. In the intervals of $0.007 \leq \Delta^* \leq 0.23$ and $0.37 \leq \Delta^* \leq 0.43$ the nematic-smectic C transition takes place, while the nematic-smectic A transition re-entrants in the intermediate range of $0.23 \leq \Delta^* \leq 0.37$. Above $\Delta^* = 0.43$ the nematic-smectic A transition is found to always be stable since the particle's shape approaches the cylinder limit with no terminal rods ($\Delta^* = 1$). In the entire range of cylinder length, $0 \leq \Delta^* \leq 1$, the nematic-columnar is never found to be stable.

The stabilization of the smectic C phase has simple geometrical reasons, since the presence of the central cylinder with non-zero length gives rise to an additional excluded volume cost in the packing entropy. As can be seen in Eq. (10) the excluded area of two central cylinders is four times that of the excluded area between the central cylinder and the terminal rod. To minimize the excluded volume cost it is favorable to form a layered structure where the central rods collide more often with the terminal rods than with each other in the layer. This can be achieved in the tilted smectic C structure where the centers of the neighboring particles are shifted relative to each other in the layer. We can also make a rough prediction for the tilt angle with the help of excluded area (Eq. 10). Assuming that two triblock particles are in contact, the smectic phase is untilted ($\alpha = 0$) when the centres of the bodies are at the same position along the z-axis with $A_{exc} = \pi D^2$. Increasing the distance between the centres along the z-axis up to $z_{12} = \Delta$, the excluded area suddenly drops to $A_{exc} = \pi D^2 / 4$ giving $\alpha \approx \arctan \frac{2\Delta}{D}$ as a geometrical expression for the tilt angle corresponding to densest packing of the triblocks. Note that our geometrical prediction for the tilt angle agrees with Eq. (15) as far as the diameter dependence is concerned. However, geometry alone is not enough to account for the density dependence of the tilt angle. The lower panel of Fig. 5 shows that the numerically obtained tilt angle and the geometrical expression show the same trend as a function of the cylinder's length and that $\tan \alpha \approx \frac{2\Delta}{D}$ can be considered as an underestimation of the transition tilt angle. The bulk properties of the smectic C phase are determined for a cylinder length of $\Delta^* = 0.02$ for various diameters and presented in Figure 6. The tilt angle, wave number and free energy are shown as a function of reduced density starting from the nematic-smectic C bifurcation point. It can be seen that for all cases the smectic C phase becomes less tilted with increasing density of the system. As shown in Figure 5 the tilt angle at the bifurcation density is greater than angle of greatest density, given by $\tan \alpha \approx \frac{2\Delta}{D}$ thus explaining the density dependence of the tilt angle. At a given density it can be seen that tilt angle increases with decreasing diameter according to Eq. (15), in agreement with the geometrical argument ($\tan \alpha \approx \frac{2\Delta}{D}$). The wave number shows a very weak, but peculiar density dependence as it can increase or decrease with density. For low values of $D^*$ it is favorable to increase the distance between the smectic layers as the triblock particle has small volume, while the trend is opposite at $D^*=1$ due to high particle's volume and the neighboring layers move closer to each other with increasing density. As the free energy does not have direct diameter dependence (Eq. 15), only two free energy curves can be seen in Fig. 6. Nematic and smectic C phases have identical free energy at the bifurcation point and the smectic C free energy is always lower than that of the nematic phase with increasing density.

Our calculations show that the smectic C phase can be stabilized in the system of parallel triblock particles. In contradiction to previous speculations, smectic C formation does not require hard particles to have a biaxial shape, such as the zig-zag and oblique cylinders. What are the implications of our work for experiment? As far as designing colloids that would exhibit the smectic C phase we recommend thin central cylindrical units and long rigid terminal units as a likely candidate as the packing fraction of the nematic-smectic C phase transition can be very low as exemplified by the relation $\eta = 2\Delta^* c$. Low volume fractions are desirable because if the transition density is very low it is improbable that the transition is preempted by another first order phase transition, such the nematic-columnar and nematic-crystal transition. Furthermore, low volume fractions mean that the kinetics of phase separation are rapid and that a gel phase is more likely to be avoided. For example, our estimate for the packing fraction of nematic-smectic C transition is 0.15 at $\Delta^* = 0.025$, which corresponds to a dilute system and, furthermore, the system forms a nematic phase if the terminal rods are enough long.

The main virture of the second virial theory is its simplicity and the fact that such theories, which have been used in the past to describe the nematic-smectic A transition of rod-like particles are in accordance with simulations. Drawbacks of the theory are that the orientational freedom of the particles and the contribution of higher virial coefficients are not included into the theory. The first point is not really a problem as the nematic phase is usually very ordered close to the nematic-smectic transition. While the higher virial coefficients can have a significant impact on the stability of the phases, inclusion of these terms would substantially increase the computational burden. To justify the prediction of the theory presented here simulation studies would be very beneficial.

## Acknowledgements

SV would like to thank the Hungarian Scientific Research Fund (grant numbers: OTKA F 47312) for financial support and SV is grateful to Kelly Services for making possible a short visit to Brandies University. SF acknowledges support from the National Science Foundation (DMR-0444172).

## Figures

**Figure 1)**

(a) Hard body representation of a linear rigid triblock particle. Top and bottom segments are rods of zero length, while the centre segment is hard cylinder with finite diameter (*D*) and length (*Δ*). The overall length of the particle is denoted by *L*. (b) Schematic of the smectic – A phase. (c) Schematic of the smectic – C phase with α

**Figure 2)**

Nematic-smectic bifurcation in the system of triblock hard particles having a central cylinder of zero cylinder length ($\Delta^* = 0$). The bifurcation density as a function of tilt angle for cylinder diameters of $D^* = 0.1$ (continuous curve), $D^* = 0.5$ (short dashed curve) and $D^* = 1$ (long dashed curve). The density, diameter and lengths are in dimensionless units: $c = \rho B_2^{ref}$, $D^* = D/L$, and $\Delta^* = \Delta/L$.

**Figure 3)**

Effect of varying length of the central cylinder on the nematic-smectic bifurcation of the system of triblock hard particles at $D^* = 1$. The bifurcation density as a function of tilt angle for a cylinder length of $\Delta^* = 0.007$ (continuous curve), $\Delta^* = 0.008$ (short dashed curve) and $\Delta^* = 0.01$ (long dashed curve). The diamond symbol shows the lowest value of the bifurcation density along each curve.

**Figure 4)**

Effect of varying diameter on the nematic-smectic bifurcation of the system of triblock hard particles at $\Delta^* = 0.02$. The bifurcation density as a function of tilt angle for various cylinder diameters; $D^* = 1$ (continuous curve), $D^* = 0.5$ (short dashed curve) and $D^* = 0.1$ (long dashed curve).

**Figure 5)**

Nematic to smectic-A, nematic to smectic-C and nematic to columnar bifurcations of the system of hard triblock particles for $D^* = 1$. (a) The bifurcation concentration as a function of cylinder length (both in dimensionless units). The continuous curve corresponds to the nematic to smectic-C bifurcation ($\alpha_{bif} > 0$), the short-dashed curve shows the nematic to smectic-A bifurcation ($\alpha_{bif} = 0$), while the long-dashed curve is the nematic-columnar bifurcation ($\alpha_{bif} = 90°$). (b) The tilt angle at the nematic to smectic-C bifurcation as a function of dimensionless cylinder length is plotted with a solid curve. The geometrical prediction for the tilt angle of the smectic C phase ($\tan\alpha \approx \frac{2\Delta}{D}$) is denoted by a dashed curve.

**Figure 6)**

Bulk properties of the smectic C phase in the tilt angle-density (a), wave number-density (b) and free energy density-density planes (c). The length of the central cylinder is the same in all cases ($\Delta^* = 0.02$), while the diameter of the central cylinder ($D^*$) equals 1 (continuous curve), 0.5 (short dashed curve) and 0.02 (dashed curve). The inset of panel (b) shows the density dependence of the wave number at a higher resolution for $D^* = 1$. In (c), the free energy of the nematic phase is denoted by a dashed curve, while that of the smectic C phase is continuous. The density, wave number and the free energy density are in dimensionless unit: $c = \rho\frac{\pi L D^2}{8}$ , $k^* = kL$, and $f^* = \beta F B_2^{ref}/V$.

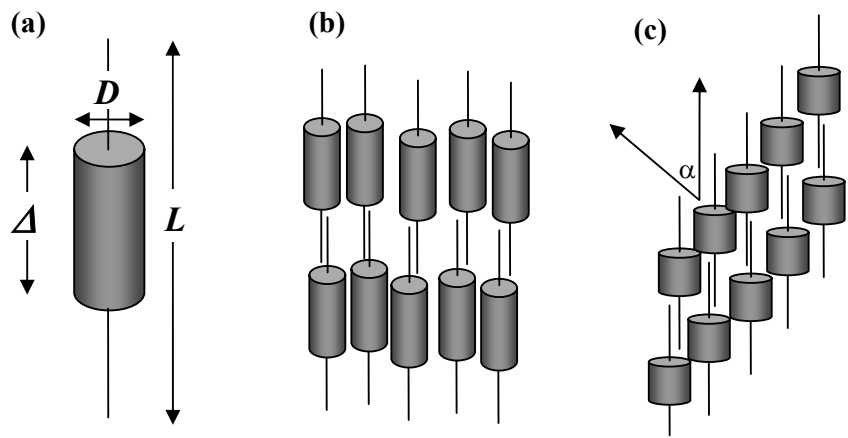

**Figure 1.**

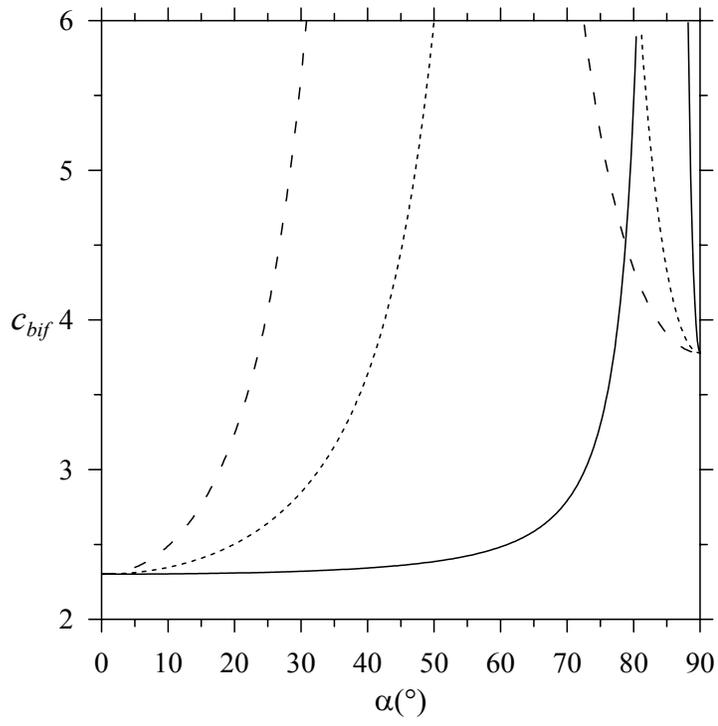

**Figure 2.**

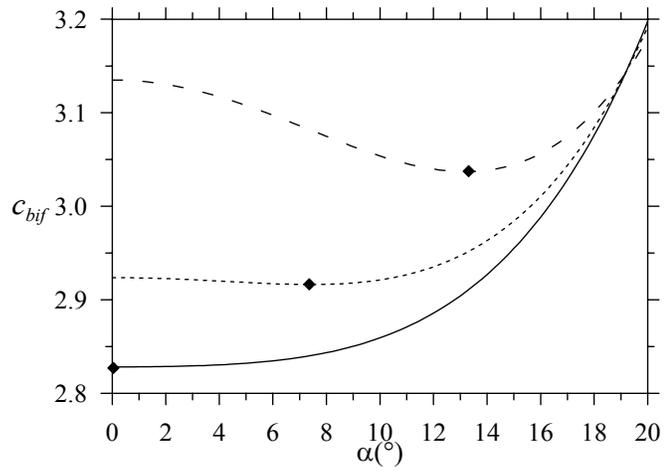

**Figure 3.**

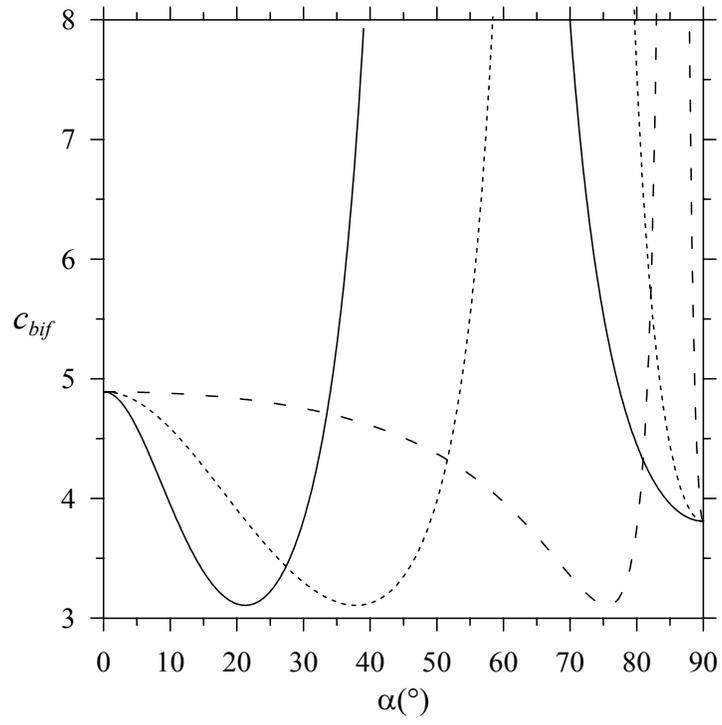

**Figure 4**

**(a)**

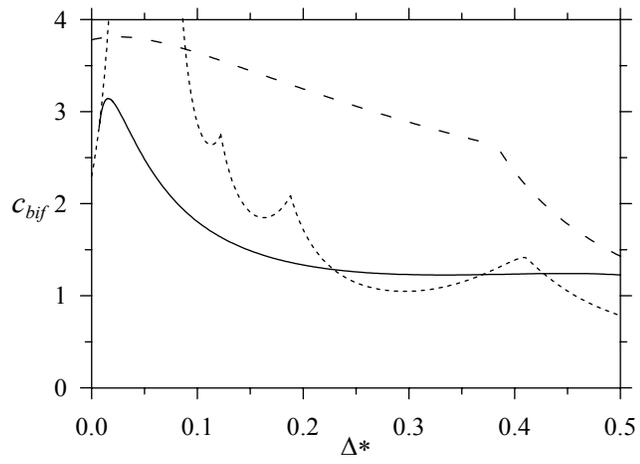

**(b)**

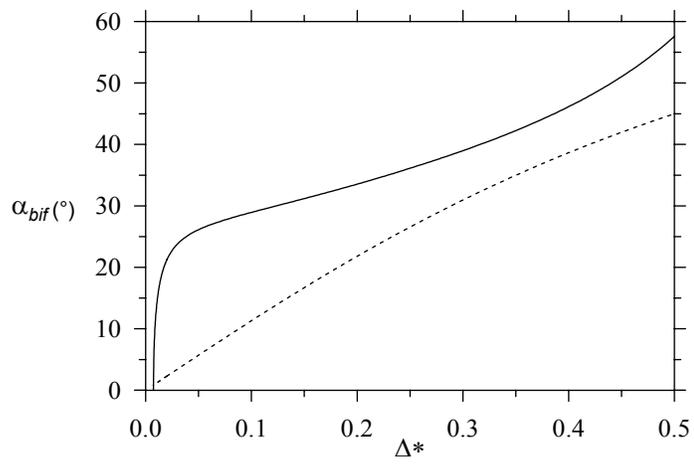

**Figure 5.**

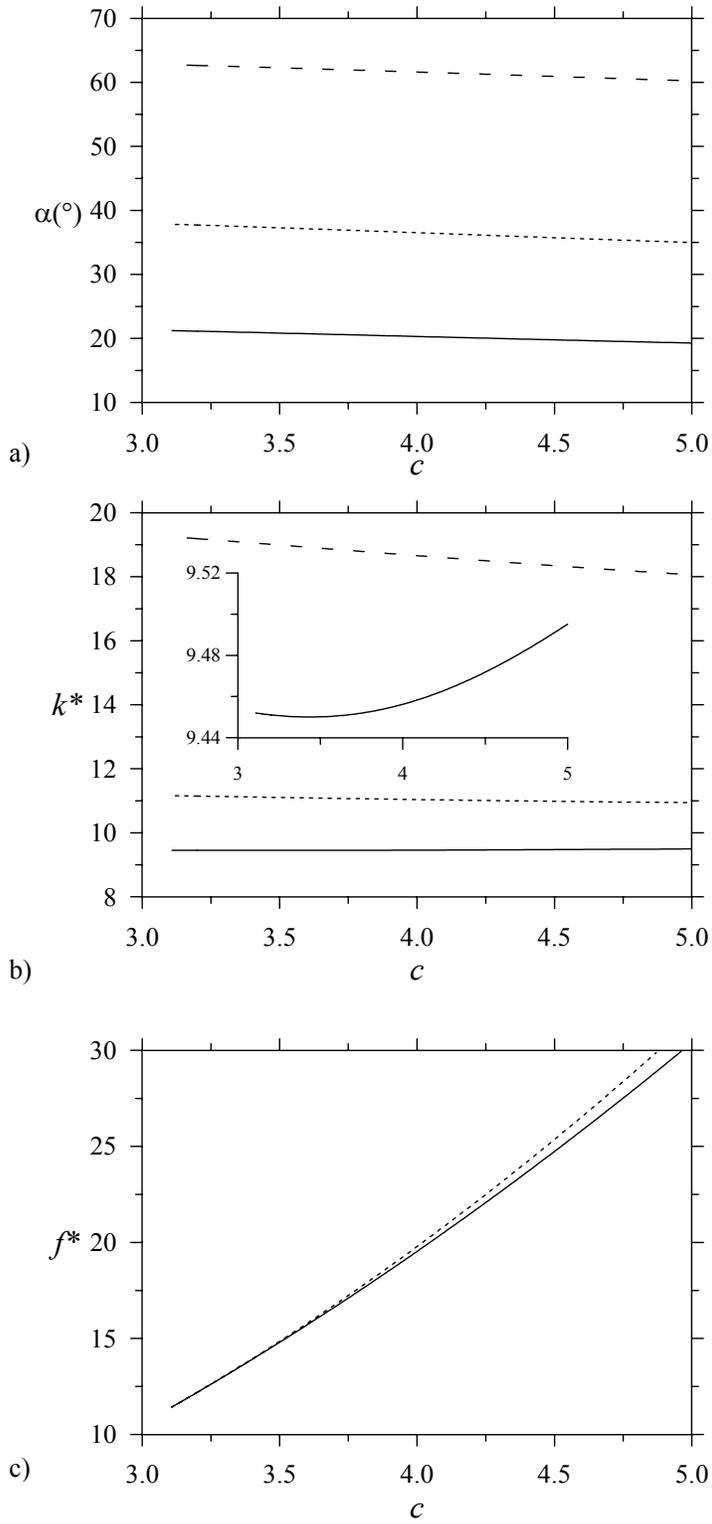

**Figure 6.**